\documentclass[twocolumn,superscriptaddress,aps,prb]{revtex4-1}

\usepackage{bm}
\usepackage{verbatim}
\usepackage{appendix}
\usepackage{graphicx}
\usepackage{amsmath}
\usepackage{amssymb}%
\usepackage{color}
\usepackage{lipsum}
\usepackage{multirow}
\usepackage[colorlinks=true,citecolor=blue,urlcolor=blue]{hyperref}
\usepackage{mathrsfs}
\usepackage{braket}
\usepackage{hhline}

\begin{document}

\title{Topologically protected boundary discrete time crystal for a solvable model}
\author{Peng Xu}
\affiliation{School of Physics, Zhengzhou University, Zhengzhou 450001, China}
\author{Tian-Shu Deng}
\email{tianshudeng500@gmail.com}
\affiliation{Institute for Advanced Study, Tsinghua University, Beijing,100084, China}

\begin{abstract}{Floquet time crystal, which breaks discrete time-translation symmetry, is an intriguing phenomenon in non-equilibrium systems. It is crucial to understand the rigidity and robustness of discrete time crystal (DTC) phases in a many-body system, and finding a precisely solvable model can pave a way for understanding of the DTC phase. Here, we propose and study a solvable spin chain model by mapping it to a Floquet superconductor through the Jordan-Wigner transformation. The phase diagrams of Floquet topological systems are characterized by topological invariants and tell the existence of anomalous edge states. The sub-harmonic oscillation, which is the typical signal of the DTC, can be generated from such edge states and protected by topology. We also examine the robustness of the DTC by adding symmetry-preserving and symmetry-breaking perturbations. Our results on topologically protected DTC can provide a deep understanding of the DTC when generalized to other interacting or dissipative systems.}
\end{abstract}

\maketitle


\section{Introduction}
Periodic driving is a powerful tool to engineer variety of unique and fascinating phenomena in many-body systems, such as Mott-insulator-superfluid transition\cite{Mott1}, dynamical gauge field\cite{DynGF1,DynGF2}, many-body echo\cite{mbe1, mbe2, mbe3}, the
realization of topological band structures\cite{ExpFloTopo1,ExpFloTopo2}. Recently, Floquet systems also bring new possibilities to simulate time-translation symmetry broken phases, which are called discrete time crystals (DTC), and this has attracted much attention from both experimentalists \cite{DTCexp1,DTCexp2,DTCexp3,DTCexp4,DTCexp5,DTCexp6,DTCexp7,DTCexp8,DTCexp9,DTCexp10,DTCexp11,DTCexp12} and theorists\cite{DTCthe1,DTCthe2,DTCthe3,DTCthe4,DTCthe5,DTCthe6,DTCthe7,DTCthe8}. 
The concept of the time crystal was originally proposed by Wilczek\cite{Wilczek}. However, it was ruled out by the no-go theorem in thermal equilibrium systems\cite{nogo1,nogo2}. Then Shivaji Sondhi's~\cite{Phase2016Sondhi} and Chetan Nayak's~\cite{Floquet2016Nayak} groups generalized the concept of time crystal and proposed the DTC, which exhibits a unique property that the expectation values of generic observables manifest a sub-harmonic oscillation. For example, the kicked Ising chain model with disordered interaction, where spins collectively flip after one period and back to their initial state after two periods, is a canonical realization of the DTC. 

In a non-interacting spin chain system, taking $\hat{U}=\exp{(-i\theta\sum_j\hat{X}_j)}$ with $\theta=\pi/2$ as a Floquet evolution operator is a straightforward method to flip all spins in one period. Here, $\hat{X}_j$ is a spin operator acting on site $j$. However, when $\theta$ deviates slightly from $\pi/2$, the period of observables also deviates from twice of the Floquet period. It means the sub-harmonic response induced by $\hat{U}=\exp{(-i\theta\sum_j\hat{X}_j)}$ is a fine-tuned result and easily destroyed by perturbations. This simple example implies that the robustness of sub-harmonic response is a crucial property of the DTC. According to the previous results, many-body localization and pre-thermalization may provide two mechanisms to stabilize the sub-harmonic response. Besides, topologically protected anomalous edge states \cite{FloEdge1,FloEdge2,FloEdge3} also suggest another mechanism of generating the robust DTC and the reasons are listed in the following. Firstly, edge states in topological insulators (superconductors) are protected by symmetries. As long as the symmetries are not broken and the gap does not close, topological edge states are stable and robust. Secondly, Floquet topological insulators (superconductors) host anomalous edge states with quasi-energy $\pi/T$ which can generate a sub-harmonic response to driving frequency $2\pi/T$. Although the relation between $\pi$-mode edge states in Floquet topological system and the DTC has been discussed in previous research\cite{TopoDTC1,TopoDTC2,TopoDTC3,TopoDTC4,TopoDTC5,TopoDTC6}, detailed and systematic of topologically protected DTC in a more general spin chain model are still lacking. Bridging Floquet topological superconductors to the topologically protected DTC in a general periodic driven spin chain model and explicitly analyzing the dynamics of observables are significantly helpful for deeply understanding the existence and robustness of the DTC.

In this work, we investigate the existence of the DTC phase in a general Floquet spin chain model. Such a periodically driven spin chain can be mapped to a Floquet superconductor through the Jordan-Wigner transformation, after which the model becomes the form of Majorana fermion. This model is intrinsic with particle-hole symmetry. Furthermore, this system can be classified into D class or BDI class, which is dependent on whether chiral symmetry is preserved. The topological Floquet superconductor exhibit a special kind of topologically non-trivial phase, where anomalous edge states with quasi-energy $\pi / T$ exist. In order to observe a robust DTC, the observable should be selected as the anomalous edge mode or the end spin. We numerically demonstrate that both observables manifest a sub-harmonic response with a generic initial product state. Finally, we also confirm the robustness of the DTC by adding symmetry-preserving and symmetry-breaking perturbations. 

The paper is organized as follows. In section \ref{sec:model}, we describe a general periodically driven spin chain model and map it to a Floquet superconductor. In section \ref{sec:topoClass}, we discuss the topological classification of this model, calculate the topological invariants and obtain the phase diagrams. In section \ref{sec:TC}, we demonstrate the existence of the DTC resulting from the Floquet superconductors, by selecting the observable as the anomalous edge mode or the end spin. Finally, in section \ref{sec:Robust}, we examine the robustness of the DTC is protected by topologically non-trivial phase by adding symmetry-preserving and symmetry-breaking perturbations.

\section{Model: Periodically Driven Spin Chain}
\label{sec:model}

\begin{figure}
\includegraphics[width=8cm]{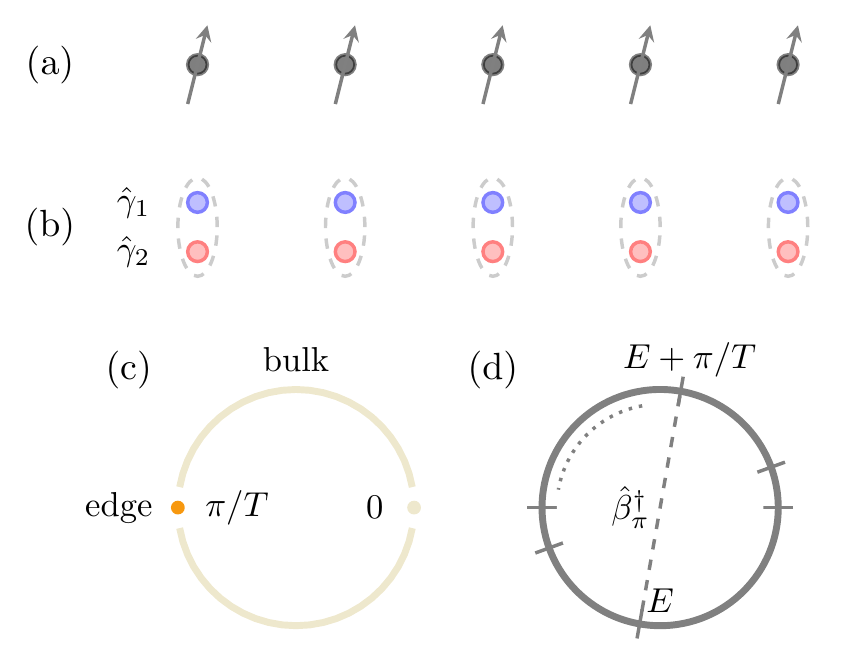}
\caption{(a) Schematic of spin chain. (b) Schematic of Majorana chain. (c) Quasi-energy spectrum $\exp(-i\epsilon_{\mu}T)$ with $\epsilon_\mu$ the quasi-energy excitation. $0$ and $\pi/T$ represent two edge states quasi-energy excitation. (d) Eigenvalues of the Floquet evolution operator $\hat{U}_T$.  Quasi-particle operator $\hat{\beta}^\dagger_\pi$ excites a state $|E\rangle$ with eigenenergy $E$ to another state $|E + \pi / T\rangle$.}
\label{fig:Setup}
\end{figure}

We consider a periodically driven spin-$1/2$ chain
as illustrated in Fig. \ref{fig:Setup} (a),
\begin{eqnarray}
\label{Ht12}
\hat{H}(t)=\begin{cases}
\hat{H}_{1}, & \text{for}\quad nT\leqslant t<nT+t_{1}\\
\hat{H}_{2}, & \text{for}\quad nT+t_{1}\leqslant t<(n+1)T
\end{cases},
\end{eqnarray}
where
\begin{equation}
\begin{aligned}
\label{HiSpin}
&\hat{H}_{m}=J_{m}^{xx}\sum_{j}\hat{X}_{j}\hat{X}_{j+1}+J_{m}^{yy}\sum_{j}\hat{Y}_{j}\hat{Y}_{j+1}\\
&+J_{m}^{xy}\sum_{j}\hat{X}{}_{j}\hat{Y}_{j+1}+J_{m}^{yx}\sum_{j}\hat{Y}_{j}\hat{X}_{j+1}+h_{m}^{z}\sum_{j}\hat{Z}{}_{j}.
\end{aligned}
\end{equation}
$\hat{X}_j$, $\hat{Y}_j$ and $\hat{Z}_j$ are spin operators (with the form of Pauli matrices) acting on the $j$-th site  and the chain contains $N$ spins. $J_m^{xx}$, $J_m^{yy}$, $J_m^{xy}$, $J_m^{yx}$ represent the strengths of nearest spin interactions and $h_m^z$ the transverse field during $m$-th time interval. By employing the Jordan-Wigner transformation
\begin{equation}
\begin{aligned}
\hat{X}_{j}&=(\hat{c}_{j}^{\dagger}+\hat{c}_{j})e^{i\pi\sum_{l<j}\hat{c}_{l}^{\dagger}\hat{c}_{l}},\\
\hat{Y}_{j}&=-i(\hat{c}_{j}^{\dagger}-\hat{c}_{j})e^{i\pi\sum_{l<j}\hat{c}_{l}^{\dagger}\hat{c}_{l}},\\
\hat{Z}_{j}&=2\hat{c}_{j}^{\dagger}\hat{c}_{j}-1.
\end{aligned}
\end{equation}
where $\hat{c}_j^\dagger$ and $\hat{c}_j$ are the creation and annihilation fermion operators on $j$-th site. Then the Hamiltonian in Eq.~(\ref{HiSpin}) can be mapped to a fermionic system,
\begin{equation}
\begin{aligned}
\label{Hcc}
\hat{H}_{m}&=\sum_{j}\left(J_{m}^{xx}-J_{m}^{yy}-iJ_{m}^{xy}-iJ_{m}^{yx}\right)\hat{c}_{j}^{\dagger}\hat{c}_{j+1}^{\dagger}\\
&+\sum_{j}\left(J_{m}^{xx}+J_{m}^{yy}+iJ_{m}^{xy}-iJ_{m}^{yx}\right)\hat{c}_{j}^{\dagger}\hat{c}_{j+1}\\
&+\frac{h_{m}^{z}}{2}\sum_{j}(2\hat{c}_{j}^{\dagger}\hat{c}_{j}-1)+h.c..\\
\end{aligned}
\end{equation}

Furthermore, by defining $\hat{\gamma}_{j, 1}=\frac{1}{\sqrt{2}}\left(\hat{c}_{j}^{\dagger}+\hat{c}_{j}\right)$ and $\hat{\gamma}_{j, 2}=i \frac{1}{\sqrt{2}}\left(\hat{c}_{j}-\hat{c}_{j}^{\dagger}\right)$, we obtain a Hamiltonian in Majorana representaion, as illustrated in Fig. \ref{fig:Setup} (b),
\begin{equation}
\begin{aligned}
\label{Hgg}
\hat{H}_m&= 2 J_{m}^{x x} \sum_{j} i \hat{\gamma}_{j, 2} \hat{\gamma}_{j+1,1}-2 J_{m}^{y y} \sum_{j}i \hat{\gamma}_{j, 1} \hat{\gamma}_{j+1,2} \\
&+2 J_{m}^{x y} \sum_{j} i \hat{\gamma}_{j, 2} \hat{\gamma}_{j+1,2}-2 J_{m}^{y x} \sum_{j}i \hat{\gamma}_{j, 1} \hat{\gamma}_{j+1,1} \\
&+2 J_{m}^{z} \sum_{j} i \hat{\gamma}_{j, 2} \hat{\gamma}_{j, 1}. \\
\end{aligned}
\end{equation}
Both Hamiltonians $\hat{H}_1$ and $\hat{H}_2$ in Eq. (\ref{Hgg}) are quadratic, so they can be rewritten as
\begin{equation}
\hat{H}_m=\frac{1}{2}\hat{\Psi}^\dagger H_m \hat{\Psi},
\end{equation}
with 
\begin{equation}
\label{Psigamma}
\hat{\Psi}^{\dagger}=\left(\begin{array}{ccccc}
\hat{\gamma}_{1,1} & \hat{\gamma}_{1,2} & \hat{\gamma}_{2,1} & \hat{\gamma}_{2,2} & ...\end{array}\right),
\end{equation}
where $H_m$ is a $2N*2N$ anti-symmetric matrix. Here and after, $\hat{H}$ ($\hat{U}$) represents an operator and $H$ ($U$) represents a matrix. Applying Fourier transformation $\hat{\gamma}_{j,m}=\frac{1}{\sqrt{N}}\sum_{k}\hat{\gamma}_{k,m}e^{ikj}$, $\hat{H}_m(k)$ has such a form 
\begin{align}
\hat{H}_m=
&\sum_{k}\left(\begin{array}{cc}
\hat{\gamma}_{k,1}^\dagger & \hat{\gamma}_{k,2}^\dagger \end{array}\right)
\left[d^{0}_m(k)I+\bm{d}_m(k)\cdot\bm{\sigma}\right]
\left(\begin{array}{c}
\hat{\gamma}_{k,1}\\
\hat{\gamma}_{k,2}
\end{array}\right),
\end{align}
where $\bm{\sigma} = (\sigma_x,\sigma_y,\sigma_z)$ denote Pauli matrices. $d^0_m(k)$ and $\bm{d}_m(k)$ are given by
\begin{equation}
\begin{aligned}
d_{m}^{0}(k) &=\left(J_{m}^{y x}-J_{m}^{x y}\right) \sin (k), \\
d_{m}^{x}(k) &=\left(J_{m}^{y y}-J_{m}^{x x}\right) \sin (k), \\
d_{m}^{y}(k) &=J_{m}^{z}+\left(J_{m}^{x x}+J_{m}^{y y}\right)\cos (k), \\
d_{m}^{z}(k) &=\left(J_{m}^{x y}+J_{m}^{y x}\right) \sin (k).
\end{aligned}
\end{equation}

For the periodically driven Hamiltonian in Eq.~(\ref{Ht12}), the Floquet evolution operator $\hat{U}_T$ and the effective Hamiltonian $\hat{H}_{\text{eff}}$ are defined by
\begin{align}
\label{expHeff}
\hat{U}_{T}&=\hat{\mathcal{T}}\exp\left(-i\int_{0}^{T}\hat{H}(t)dt\right)=\exp(-i\hat{H}_{2}t_{2})\exp(-i\hat{H}_{1}t_{1})\nonumber\\
&=\exp(-i\hat{H}_{\rm{eff}}T),
\end{align}
where $t_2+t_1=T$. Based on the quadratic form of $\hat{H}_1$ and $\hat{H}_2$ in Eq.~(\ref{Hgg}), we have proven in Appendix \ref{app:a} that the Floquet effective Hamiltonian is also quadratic, which means $\hat{H}_{\rm{eff}}$ also takes the form
\begin{equation}
\hat{H}_{\rm{eff}}= \frac{1}{2}\hat{\Psi}^{\dagger} H_{\rm{eff}} \hat{\Psi}.
\end{equation}
In order to obtain the eigenenergies of $\hat{H}_{\rm{eff}}$, we diagonalize the above matrix $H_{\rm{eff}}$ in real space as
\begin{equation}
	H_{{\rm eff}}=V\Lambda^{\epsilon}V^{\dagger},
\end{equation}
where
\begin{equation}
	V=\left(\begin{array}{cccc}
	|V_{1}\rangle & |V_{2}\rangle & ... & |V_{2N}\rangle\end{array}\right),
\end{equation}
\begin{equation}
\Lambda=\left(\begin{array}{cccc}
\epsilon_{1}\\
 & \epsilon_{2}\\
 &  & \ddots\\
 &  &  & \epsilon_{2N}
\end{array}\right).
\end{equation}
$|V_{\mu}\rangle$ is the eigenvector of $H_{\rm{eff}}$ with eigenvalue $\epsilon_{\mu}$ such that 
$H_{{\rm eff}}|V_{\mu}\rangle=\epsilon_{\mu}|V_{\mu}\rangle$. $\epsilon_{\mu}$ is the Bogoliubov quasi-particle excitation spectrum of $\hat{H}_{\rm{eff}}$, which is restricted to the Floquet Brillouin Zone $[-\pi/T,\pi/T)$. Then, the Floquet effective Hamiltonian $\hat{H}_{\rm{eff}}$ in this quasi-particle representation can be rewritten as
\begin{equation}
	\label{Heffhat}
\hat{H}_{{\rm eff}}=\frac{1}{2}\hat{\Psi}^{\dagger}V\Lambda^{\epsilon}V^{\dagger}\hat{\Psi}=\frac{1}{2}\hat{\Phi}^{\dagger}\Lambda^{\epsilon}\hat{\Phi},
\end{equation} 
where $\hat{\Phi}$ is composed of Bogoliubov quasi-particles $\hat{\Phi}^{\dagger}=\left(\begin{array}{cccc}
	\hat{\alpha}_{1}^{\dagger} & \hat{\alpha}_{2}^{\dagger} & \cdots & \hat{\alpha}_{2N}^{\dagger}\end{array}\right)$ and $\hat{\alpha}^\dagger_\mu=\hat{\Psi}^\dagger|V_\mu\rangle$.

Due to the particle-hole symmetry constraint, the eigenvalues and eigenvectors of $H_{\rm{eff}}$ must come in pair
\begin{equation}
\begin{aligned}
	\label{UFV}
	H_{\rm{eff}}|V_\mu\rangle&=\epsilon_\mu|V_\mu\rangle,\\
	H_{\rm{eff}}|V_\mu^\ast\rangle&=-\epsilon_\mu|V_\mu^\ast\rangle.
\end{aligned}
\end{equation}
Substituting these relations into Eq.~({\ref{Heffhat}}), the effective Hamiltonian $\hat{H}_{\rm{eff}}$ becomes 
\begin{equation}
\label{HatHeff}
\hat{H}_{{\rm eff}}=\sum_{\epsilon_{\mu}>0}\epsilon_{\mu}(\hat{\alpha}_{\mu}^{\dagger}\hat{\alpha}_{\mu}-\frac{1}{2}).
\end{equation}
The eigenstate of $\hat{H}_{\rm{eff}}$ is given by $\prod_{\mu}|n_{\mu}\rangle=(\alpha_{\mu}^{\dagger})^{n_{\mu}}|0\rangle$.
$n_\mu=0,1$ represents the number of occupations on quasi-particle mode. Then the energy spectrum of this system can be expressed as
\begin{align}
    E&=\langle\hat{H}_{{\rm eff}}\rangle\nonumber\\
    &=\sum_{\epsilon_{\mu}>0}\epsilon_{\mu}(\langle\alpha_{\mu}^{\dagger}\alpha_{\mu}\rangle-\frac{1}{2})\nonumber\\
    &=E_0+\sum_{\mu}\epsilon_\mu{n}_{\mu},
\end{align}
where $E_0=-\frac{1}{2}\sum_{\epsilon_{\mu}>0}\epsilon_\mu$ is the energy of ground state and different configurations of ${n}_\mu$ generate $2^N$ eigenenergies of the spin chain system.

To arrive at our final destination of observing the DTC in this periodically driven spin chain, we firstly concentrate on the evolution of operator $\hat{\alpha}^\dagger_\mu$
\begin{align}
\label{alphamu}
\hat{\alpha}^\dagger_{\mu}(nT)&=\hat{U}_{T}^{-n}\hat{\alpha}^\dagger_{\mu}(0)\hat{U}_{T}^{n}
=\hat{\alpha}^\dagger_{\mu}(0)e^{inT\epsilon_{\mu}}.
\end{align}
For a given initial state, the expectation value of the operator $\hat{\alpha}^\dagger_{\mu}$ oscillates with period ${2\pi}/{\epsilon_\mu}$. Generally, this oscillating behavior is not rigid and easily perturbed, so the performance for operators $\hat{\alpha}^\dagger_{\mu}(nT)$ cannot be regarded as a signal of the DTC phase. However, there is an exception due to anomalous edge states in Floquet topological systems. Such systems exhibit two kinds of edge states with eigenenergies $\epsilon_\mu=0$ and $\epsilon_\mu=\pi/T$, protected by topological invariants $\nu_0$ and $\nu_\pi$, respectively. Here, we mainly focus on the edge states with eigenenergy $\epsilon_\mu=\pi/T$, denoted as $H_{\rm {eff}}|W\rangle =\pi/T|W\rangle$ and we define $\hat{\beta}^\dagger_\pi=\hat{\Psi}^\dagger|W\rangle$. The expectation value of the operator $\hat{\beta}^{\dagger}_\pi$ exactly oscillates with period $2T$ as shown in Eq.~(\ref{alphamu}). More importantly, this oscillation period is protected by the Floquet topologically non-trivial phases such that this phenomenon of the DTC is robust against symmetry-preserving perturbations.

To further demonstrate the DTC induced by Floquet topological superconductors,  we systematically discuss the phase diagrams of Floquet topological superconductors in section \ref{sec:topoClass} and  discuss the DTC for different observables, including anomalous edge operators and end spin operators in section \ref{sec:TC}.

\section{Floquet topological phases}
\label{sec:topoClass}
Following the standard approach of studying topological superconductors,  we calculate the topological invariants of the effective Hamiltonian in momentum space. $\hat{H}_{\rm{eff}}$ is expressed as
\begin{equation}
	\hat{H}_{{\rm eff}}=\sum_{k}\left(\begin{array}{cc}
		\hat{\gamma}_{k,1} & \hat{\gamma}_{k,2}\end{array}\right)
		H_{\rm{eff}}(k)
		\left(\begin{array}{c}
		\hat{\gamma}_{k,1}\\
		\hat{\gamma}_{k,2}
	\end{array}\right),
\end{equation}
where $H_{\rm{eff}}(k)$ are obtained from
\begin{equation}
\begin{aligned}
	U(k)&=\exp[-iH_{2}(k)t_{2}]\exp[-iH_{1}(k)t_{1}] \\
	&=\exp[-iH_{{\rm eff}}(k)T]
\end{aligned}	
\end{equation}
according to Appendix \ref{app:a}.
Here, $d_{{\rm eff}}^{0}(k)$ and $\boldsymbol{d}_{{\rm eff}}(k)$ are defined using $H_{{\rm eff}}(k)=d_{{\rm eff}}^{0}(k)I+\boldsymbol{d}_{{\rm eff}}(k)\cdot\boldsymbol{\sigma}$. In the following, we request $\epsilon(k)=d_{{\rm eff}}^{0}(k)\pm\sqrt{d_{{\rm eff}}^{2}(k)}\neq 0$ or $\pi / T$ for any $k\in[-\pi,\pi)$, due to the topological superconductors should be gapped.

For a one-dimensional two-band system with intrinsic particle-hole constraint, topologically non-trivial phases can only be classified into classes BDI (with chiral symmetry) and D (without chiral symmetry) \cite{foot1,classTopo}. So, in the following, we discuss the calculations of topological invariants for classes D and BDI, respectively.

\subsection{D Class}

For simplicity, we take $J_1^{yy}=J_2^{yy}=0, J_2^{xx}=h_1^z=0$, non-zero coupling strengths  $J_{1}^{xy}=J_{2}^{xy}=J_{1}^{yx}= J_{2}^{yx}$ to break chiral symmetry. Besides, here and after, we take $t_1=t_2=t$. Then $H_1(k)$ and $H_2(k)$ is given by
\begin{align}
    &H_1(k)=-J_1^{xx}(\sin(k)\sigma_x-\cos(k)\sigma_y)+(J_{1}^{xy} + J_1^{yx})\sin(k)\sigma_z,\nonumber\\
    &H_2(k)=h_2^z\sigma_y+(J_{1}^{xy} + J_1^{yx})\sin(k)\sigma_z.
\end{align}
Following the approach in Ref.\cite{FloquetTopo1}, topological invariants $\nu_0$ and $\nu_\pi$ in Floquet superconductors have such forms
\begin{equation}
\begin{aligned}
\label{Q0Qp}
\nu_{0}\nu_{\pi}&=\operatorname{sgn}\left[{\rm{Pf}}(M_{0})\right] \operatorname{sgn}\left[{\rm{Pf}}(M_{\pi})\right], \\
\nu_{0}&=\operatorname{sgn}\left[{\rm{Pf}}(N_{0})\right] \operatorname{sgn}\left[{\rm{Pf}}(N_{\pi})\right],
\end{aligned}
\end{equation}
where $\operatorname{Pf}\left[X\right]$ is the Pfaffian number of a skew marix $X$ and $M_{k}=\log\left[U(k)\right], N_{k}=\log\left[\sqrt{U(k)}\right]$. Since both $\log({X})$ and $\log(\sqrt{X})$ are multivalued functions, for $k=0$ and $k=\pi$ we have
\begin{equation}
\label{MNxik}
M_{k}=-i\xi_{M}(k)\sigma_{y},\quad
N_{k}=-i\xi_{N}(k)\sigma_{y},
\end{equation}
with
\begin{equation}
\begin{aligned}
\label{xiMNk}
\xi_{M}(k)+2z\pi&=J_1^{xx}t\cos(k)+h_2^zt,\\
\xi_{N}(k)+2z\pi&=\frac{J_1^{xx}t\cos(k)+h_2^zt}{2}.
\end{aligned}
\end{equation}
Here, $z$ is taken as an appropriate integer so that $\xi_{M/N}(k)$ can be constrained to the interval $[-\pi,\pi)$. Finally, we obtain 
\begin{equation}
\begin{aligned}
\nu_{0}\nu_{\pi}&={\rm sgn}[\sin(P_{+})\sin(P_{-})],\\
\nu_{0}&={\rm sgn}[\sin(\frac{P_{+}}{2})\sin(\frac{P_{-}}{2})].
\end{aligned}
\end{equation}
with
\begin{equation}
\begin{aligned}
P_{+}&=h_{2}^{z}t+J_{1}^{xx}t,\\
P_{-}&=h_{2}^{z}t-J_{1}^{xx}t.
\end{aligned}
\end{equation}

\begin{figure}
\includegraphics[width=8cm]{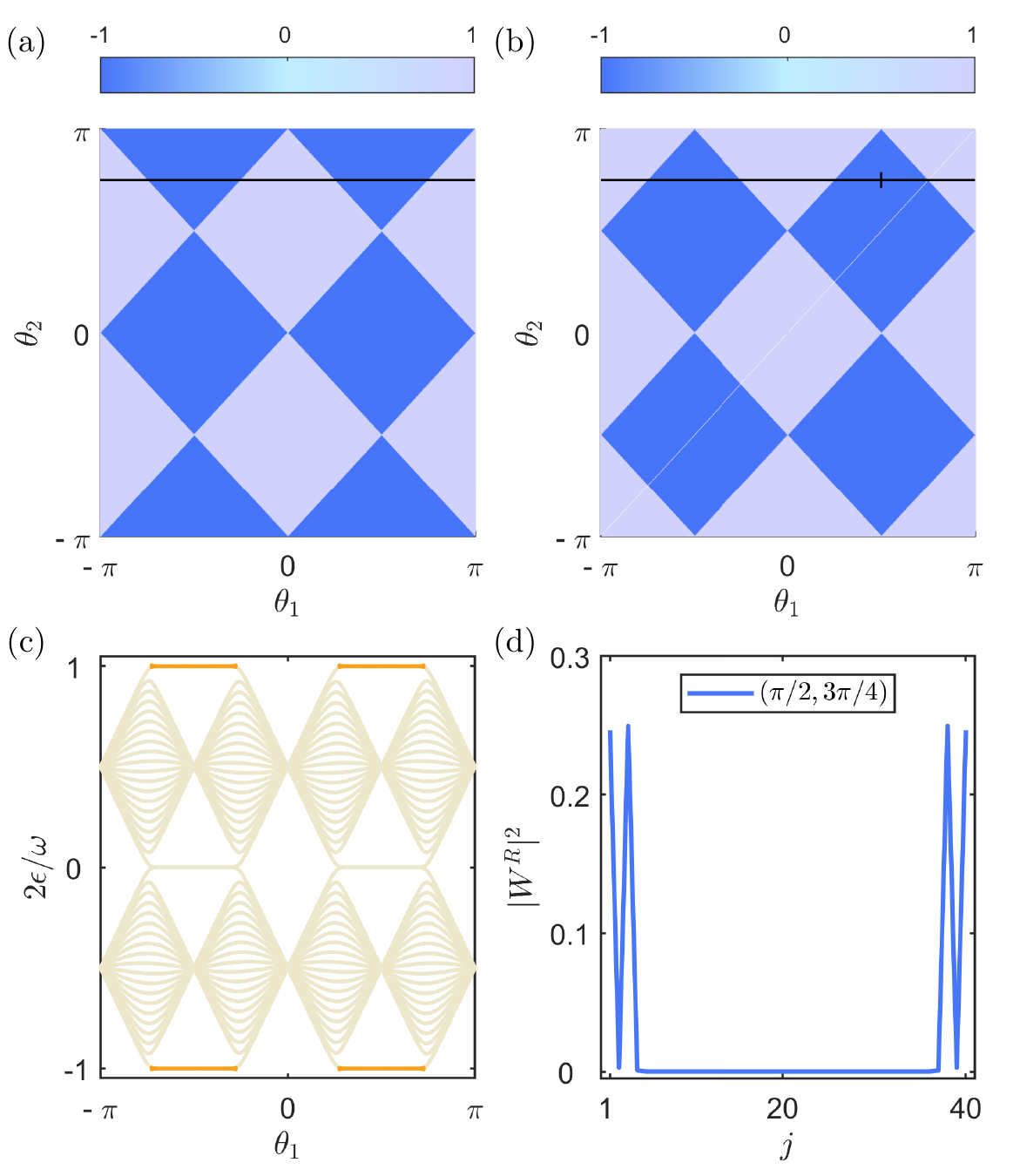}
\caption{(a, b) Phase diagrams of the D class topological superconductor (without chiral symmetry). The areas of topological invariants with value $- 1$ indicate a topologically non-trivial superconductor, otherwise topologically trivial. Topological invariant $\nu_0$ is shown in (a) and $\nu_\pi$ in (b). (c) Quasi-energy spectrum of the topological superconductor under open boundary condition with parameters $\theta_1 \in [- \pi, \pi), \theta_2 = 3 \pi / 4$ along the black solid line shown in (a, b). (d) $\pi$-mode edge state $|W^R\rangle$ with parameters $(\theta_1, \theta_2) = (\pi / 2, 3 \pi / 4)$. Here, we set $J_1^{xx} t=\theta_1, h_2^z t=\theta_2$.}
\label{fig:phase1}
\end{figure}

In Fig.~\ref{fig:phase1} (a, b), we show the phase diagrams of topological invariants $\nu_0$ and $\nu_\pi$ on the $\theta_1$-$\theta_2$ plane by taking $\theta_1=J_1^{xx}t$ and $\theta_2=h_2^{z}t$. For the  D class, the system is classified by a $\mathbb{Z}_2$ topological invariant. $\nu_0$ and $\nu_\pi$ take the value $\pm 1$, and $\nu_\pi=-1$ indicates anomalous topologically non-trivial phases. The quasi-energy spectrum as a function of $\theta_1\in[- \pi, \pi)$ is plotted in Fig.~\ref{fig:phase1} (c). The gapless $0$-mode and $\pi$-mode must exist where the topological invariants $\nu_0 = - 1$ and $\nu_\pi = - 1$, respectively. Consequently, the topologically protected discrete time crystal could be observed in the regime $\nu_\pi = - 1$.

Furthermore, because of particle-hole symmetry, the eigenvalues and eigenvectors of $H_{\rm{eff}}$ always come in pair as shown in Eq.~(\ref{UFV}). Thus for the edge states with quasi-energies $\epsilon_\mu=0$ or $\pi/T$, the eigenvectors $|V_\mu\rangle$ and $|V_\mu^*\rangle$ span a degenerate subspace, from which we can construct a pure real vector and a pure imaginary vector as $|V_\mu^R\rangle=|V_\mu\rangle+|V_\mu^*\rangle$ and $|V_\mu^I\rangle=|V_\mu\rangle-|V_\mu^*\rangle$. 
In Fig.~\ref{fig:phase1} (d), we have shown the probability distribution of an anomalous edge state $|W^{R}\rangle =(|W\rangle+|W^*\rangle)/\sqrt{2}$. For convenience, here and after, $\{j,s\}$ ($j=1,2,...L$ and $s=1,2$) in Eq. \eqref{Psigamma} is relabelled as $\{2j+s-2\}$ for the index of $\hat{\gamma}$, as well as $|W^{R}\rangle$.

\subsection{BDI Class}
For simplicity, we take non-zero coupling strengths $J_1^{xx}=J_2^{xx}$, $h_2^z$ and other parameters zero so that the system obeys chiral symmetry. With given parameters, the chiral operator is taken as $U_S=\sigma_z$ and we have
\begin{equation}
\begin{aligned}
U_S^\dagger U^{a, b} U_S = (U^{a, b})^\dagger.
\end{aligned}
\end{equation}
where 
\begin{equation}
\begin{aligned}
	{U}^{a}&=\exp(-iH_{1}t_{1}/2)\exp(-iH_{2}t_{2})\exp(-iH_{1}t_{1}/2)\\
	&=\exp({-iH_{\rm{eff}}^a} T), \\
	U^{b}&=\exp(-iH_{2}t_{2}/2)\exp(-iH_{1}t_{1})\exp(-iH_{2}t_{2}/2)\\
	&=\exp({-iH_{\rm{eff}}^b} T),
\end{aligned}
\end{equation}
and
\begin{equation}
\begin{aligned}
&H_1(k)=-J_1^{xx}(\sin(k)\sigma_x-\cos(k)\sigma_y),\\
&H_2(k)=h_2^z\sigma_y-J_1^{xx}(\sin(k)\sigma_x-\cos(k)\sigma_y).
\end{aligned}
\end{equation}
Here, we introduce $U^{a}$ and $U^b$, which are both unitarily transformed from $U(k)$, in order that the chiral operator is explicitly $\sigma_z$. Besides, based on $U^a$ and $U^b$ the topological invariants $\nu_0$ and $\nu_\pi$ can be calculated as \cite{FloquetTopo2} follows,
\begin{equation}
\nu_{0,\pi}=\frac{\nu_a\pm\nu_b}{2},
\end{equation}
where $\nu_{a}$ and $\nu_{b}$ are the winding numbers defined as
\begin{equation}
\nu_{a,b}=\frac{1}{2\pi}\int dk\frac{-d_{a,b}^{x}(k)\partial_{k}d_{a,b}^{y}(k)+d^{y}_{a,b}(k)\partial_{k}d^x_{a,b}(k)}{\left(d_{a,b}^{x}(k)\right)^{2}+\left(d_{a,b}^{y}(k)\right)^{2}},
\end{equation}
and $d_{a,b}^{x,y}(k)$ is given by
\begin{align}
H_{\rm{eff}}^{a,b}(k)&=d_{a,b}^x(k)\sigma_x+ d_{a,b}^y(k)\sigma_y.
\end{align}

\begin{figure}
\includegraphics[width=8cm]{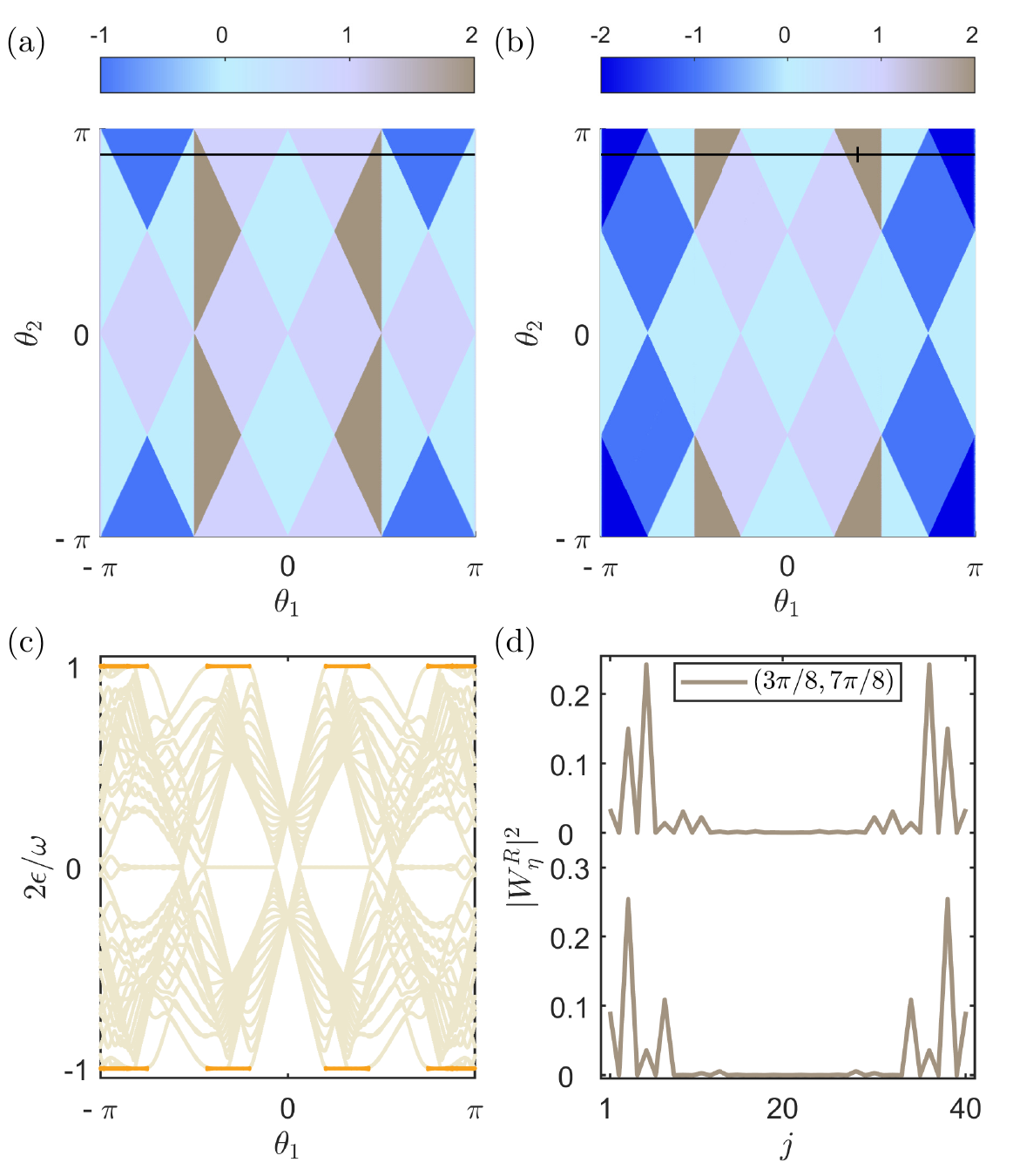}
\caption{(a, b) Phase diagrams of the BDI class topological superconductor (with chiral symmetry). The areas of topological invariants with non-zero values indicate a topologically non-trivial superconductor, otherwise topologically trivial. Topological invariant $\nu_0$ is shown in (a) and $\nu_\pi$ in (b). (c) Quasi-energy spectrum of the topological superconductor under open boundary condition with parameters $\theta_1 \in [- \pi, \pi), \theta_2 = 7 \pi / 8$ along the black solid line shown in (a, b). (d) $\pi$-mode edge state $|W^R\rangle$ with parameters $(\theta_1, \theta_2) = (3 \pi / 8, 7 \pi / 8)$. Here, we set $\theta_1 = J_1^{xx} t = J_2^{xx} t, \theta_2 = h_2^z t$.
}
\label{fig:phase2}
\end{figure}

In Fig.~\ref{fig:phase2} (a) and (b), we show the phase diagrams of topological invariants $\nu_0$ and $\nu_\pi$ on the $\theta_1$-$\theta_2$ plane. For the BDI class, the system is classified by a $\mathbb{Z}$ topological invariant. Both $\nu_0$ and $\nu_\pi$ take integer values, and $\nu_\pi\neq0$ indicates anomalous topologically non-trivial phases. The quasi-energy spectrum as a function of $\theta_1\in[- \pi, \pi)$ is plotted in Fig.~\ref{fig:phase1} (c). In contrast to D class, the topological invariant in class BDI $\nu_0$ and $\nu_\pi$ could be $\pm 2, \pm 3, \cdots $ leading to larger degeneracy of edge states. For example, for $(\theta_1,\theta_2)=(3\pi/8,7\pi/8)$, we have $\nu_\pi=2$, which results in four $\pi$-mode edge states. These four edge states are labeled as $|W_\eta\rangle$ and $|W_\eta^*\rangle$ with $\eta = 1, 2$.  In Fig.~\ref{fig:phase2} (d), we show the  the probability distribution of the anomalous edge states as $|W^R_{\eta = 1}\rangle$ and $|W^R_{\eta = 2}\rangle$, respectively.

\section{Sub-harmonic oscillations of the DTC}
\label{sec:TC}

\begin{figure}[htp]
\includegraphics[width=8cm]{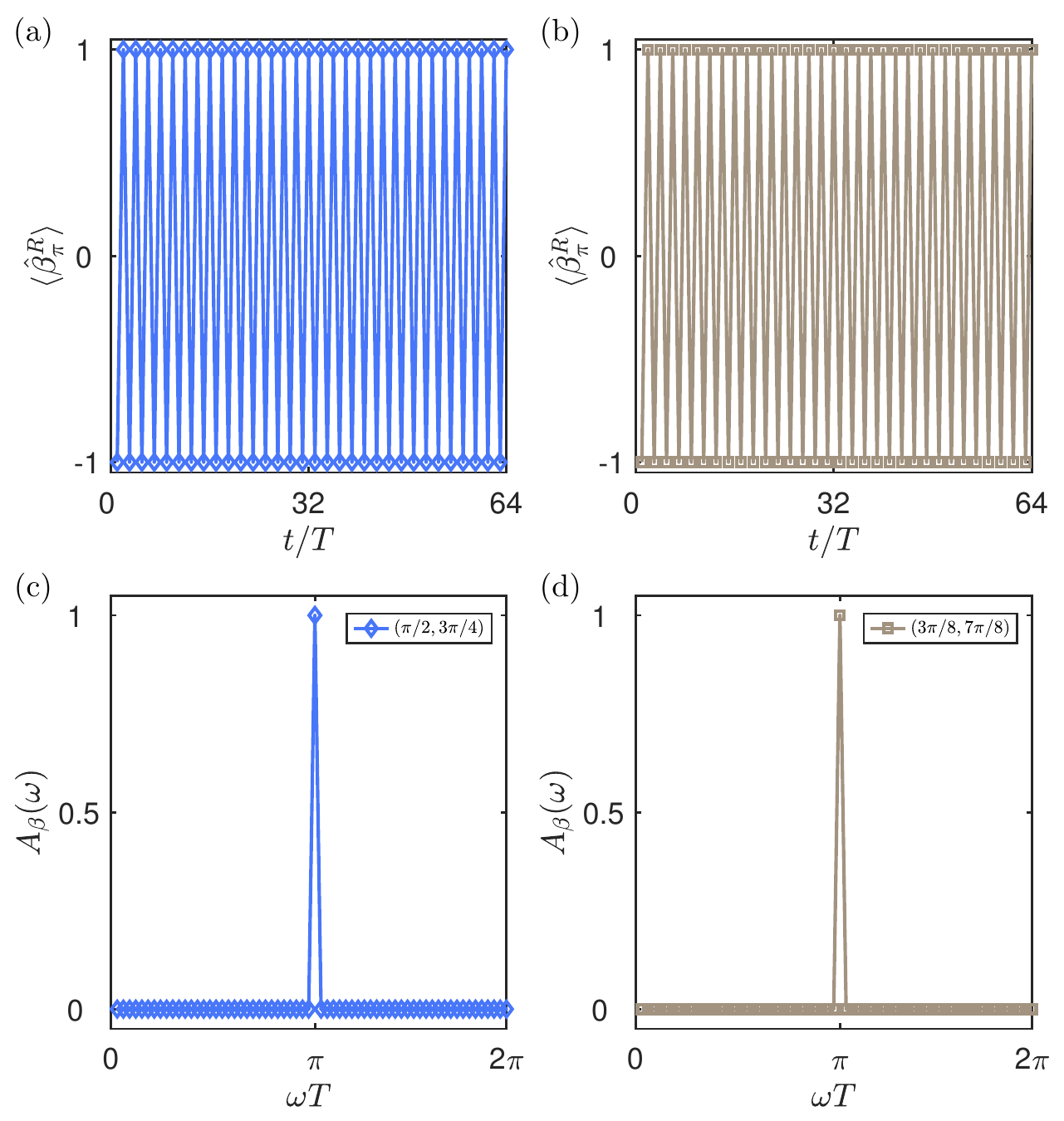}
\caption{ (a, b) Dynamics of anomalous edge mode $\hat{\beta}_\pi^R$. (c, d) Distribution of response function $A_\beta(\omega)$. The frequency distributions in (c) and (d) are the Fourier transformation of dynamics in (a) and (b), respectively. (a, c) with the same parameters as Fig.~\ref{fig:phase1} (d) correspond to D class, while (b, d) with the same parameters as Fig.~\ref{fig:phase2} (d) correspond to BDI class. The initial state is taken as a product state polarized along $x$ direction.}
\label{fig:dyn3}
\end{figure}

\begin{figure}[t]
\includegraphics[width=8cm]{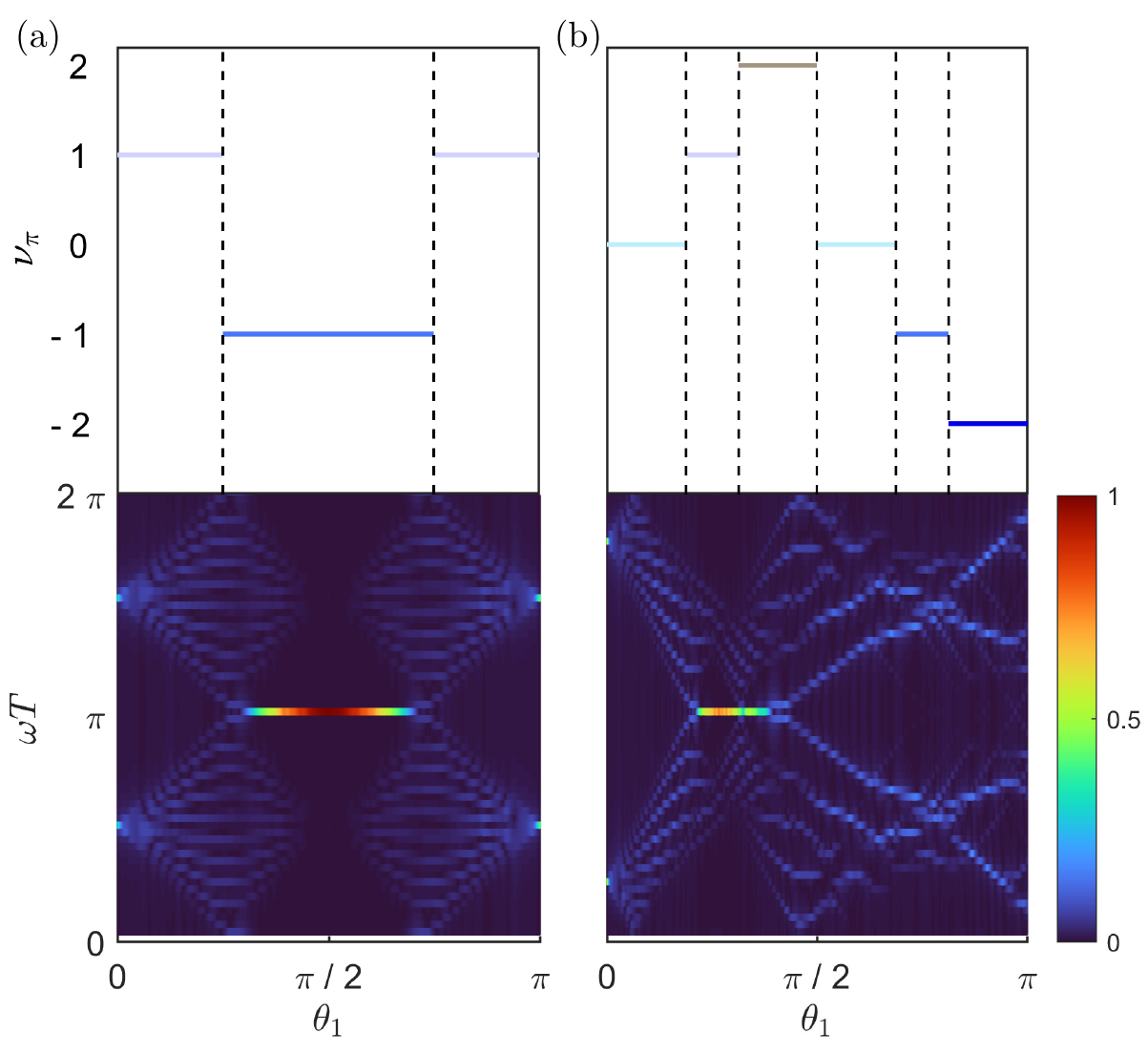}
\caption{(a, b) Upper panels: the values of topological invariants in different regimes. Lower panels: frequency distributions of the response function for observable $\hat{X}_1$ and the color represents the amplitude of $A_X(\omega)$.  Parameters of (a) are the same as the black solid line in Fig.~\ref{fig:phase1}, while parameters of (b) are the same as the black solid line in Fig.~\ref{fig:phase2}.}
\label{fig:fredis}
\end{figure}
In this section, we'll examine the sub-harmonic oscillations of the DTC in topologically non-trivial phases with $\nu_\pi=-1$ in the D class and $\nu_\pi\neq0$ in the BDI class, by selecting the anomalous edge mode and the end spin as observables.

We first take the anomalous edge mode as the observable for the DTC,
\begin{equation}
\hat{\beta}_{\pi}^{R}=\frac{\hat{\beta}_{\pi}+\hat{\beta}_{\pi}^{\dagger}}{\sqrt{2}}.
\end{equation}
As we have derived in Eq.~(\ref{alphamu}), the time evolution of observable $\hat{\beta}_\pi^R$ is expressed as
\begin{equation}
\begin{aligned}
\langle\hat{\beta}_\pi^R(nT)\rangle&=\langle\psi(0)|\hat{U}_{T}^{-n}\hat{\beta}_{\pi}^{R}\hat{U}_{T}^{n}|\psi(0)\rangle\\
&=(-1)^{n}\langle\psi(0)|\hat{\beta}_{\pi}^{R}|\psi(0)\rangle.
\end{aligned}
\end{equation}
It is obvious that the dynamics of $\langle\hat{\beta}_\pi^R(nT)\rangle$ manifests an exact sub-harmonic response for a generic initial state with $\langle\psi(0)|\hat{\beta}_{\pi}^{R}|\psi(0)\rangle\neq0$. After the Fourier transformation of $\langle\hat{\beta}_\pi^R(nT)\rangle$, we obtain the response function
\begin{equation}
\label{Oomega}
A_\beta(\omega)=\frac{1}{n_{\rm{max}}}\sum_{n=1}^{n_{\rm{max}}}\langle\hat{\beta}_\pi^R(nT)\rangle e^{i\omega nT}.
\end{equation}
where $n_{\text{max}}$ represents the maximal number of periods. In Fig.~\ref{fig:dyn3} (a) and (b), we show the expectation values of $\hat{\beta}^R_{\pi}$ as a function of evolution time for D class and BDI class, respectively. Correspondingly the response functions $A_\beta(\omega)$ are plotted  in Fig.~\ref{fig:dyn3} (c) and (d). $\langle\hat{\beta}^R_{\pi}\rangle$ oscillates with double periods and $A_\beta(\omega)$ peaks at $\omega = \pi / T$, which confirm sub-harmonic response for the observable $\hat{\beta}^R_{\pi}$ in the regime of Floquet topologically non-trivial phases with anomalous edge states. However, we have to admit that $\hat{\beta}^R_{\pi}$ is difficult to engineer and observe in experiments, and the concrete form of operator $\hat{\beta}^R_{\pi}$ in spin basis depends on coupling parameters. Therefore, it is important to choose a more appropriate operator that is easily observed in experiments.

Another observable for the DTC is taken as the end spin operator $\hat{X}_1$, since the anomalous edge states finitely occupy the end sites. Actually, the end spin $\hat{X}_1$ is nothing but $\hat{\gamma}_1$ after the Jordan-Wigner transformation, which could be decomposed as
\begin{equation}
\label{gamma1}
\hat{\gamma}_{1}=\sum_{\eta=1}^{n_{\rm{edge}}}|W_{\eta} \rangle_1\hat{\beta}_{\pi,\eta} + \sum_{\epsilon_\mu\neq\pi/T}|V_{\mu} \rangle_1\hat{\alpha}_{\mu} + h.c.,
\end{equation}
where $|W_{\eta} \rangle_1$ and $|V_{\mu} \rangle_1$ represent the first element of vectors $|W_{\eta} \rangle$ and $|V_{\mu} \rangle$, respectively. $n_{\rm edge}$ represents the number of pairs of edge states.
Dynamics of the observable $\hat{X}_{1}$ is then given by
\begin{equation}
\begin{aligned}
\langle\hat{X}_{1}(nT)\rangle&=\langle\psi(0)|\hat{U}_{T}^{-n}\hat{\gamma}_{1}\hat{U}_{T}^{n}|\psi(0)\rangle\\
&=(-1)^{n}\sum_{\eta=1}^{n_{\rm{edge}}}2{\rm{Re}}(|W_{\eta}\rangle_1\langle\psi(0)|\hat{\beta}_{\pi,\eta}|\psi(0)\rangle)\\
&+\sum_{\epsilon_{\mu}\neq\pi/T}2{\rm Re}(|V_{\mu}\rangle_1 e^{2i\epsilon_{\mu}}\langle\psi(0)|\hat{\alpha}_{\mu}|\psi(0)\rangle).
\end{aligned}
\label{eq:endspinevo}
\end{equation}
In D class, there is only one pair of edge states with $n_{\rm{edge}}=1$, which is localized around $\hat{\gamma}_1|0\rangle$ and $\hat{\gamma}_{2N}|0\rangle$. Therefore, $|W_{1}\rangle_1$ is finitely distributed so that $\langle\hat{X}_{1}(nT)\rangle$ manifests a sub-harmonic response at $\omega=\pi/T$ as is shown in the second line of Eq.~(\ref{eq:endspinevo}). In Fig.~\ref{fig:fredis} (a),
we show the response function $A_X(\omega)$ of $\hat{X}_1$, which peaks at $\omega=\pi/T$ when $\nu_\pi=-1$. These numerical results indicate the presence of the DTC for the end spin observable. But in BDI class, according to the bulk-edge correspondence\cite{asboth2016short}, $n_{\rm{edge}}=|\nu_\pi|$ and $\nu_\pi>0$ ($\nu_\pi<0$) represents the number of edge states occupying $\hat{\gamma}_{1}|0\rangle$ ($\hat{\gamma}_{2}|0\rangle$). Therefore, $|W_{\eta}\rangle_1$ is finite when $\nu_\pi>0$, while $|W_{\eta}\rangle_1$ is exactly $0$ when $\nu_\pi<0$. As shown in Fig.~\ref{fig:fredis} (b), $A(\omega)$ peaks at $\omega =\pi/T$ when $\nu_\pi>0$, which also indicats the presence of the DTC for BDI class. Note that the DTC regime, where $A(\omega)$ peaks at $\omega=\pi/T$, doesn't exactly match with the topological region with $\nu_\pi=-1$ ($\nu_\pi>0$) for D (BDI) class. We think this phenomenon is due to the finite size effect, which causes the quasi-energy of edge mode to be not perfectly $\pi/T$. In order to check the finite size effect, the numerical results for different lengths of the chain are shown in Appendix \ref{app:b}.

\section{Robustness of the DTC}
\label{sec:Robust}
In order to demonstrate the robustness of the DTC induced by topological superconductors, we continue to investigate the frequency distributions of response function for $\hat{X}_1$ after adding symmetry-preserving and symmetry-breaking perturbations.

\begin{figure}[t]
\includegraphics[width=8cm]{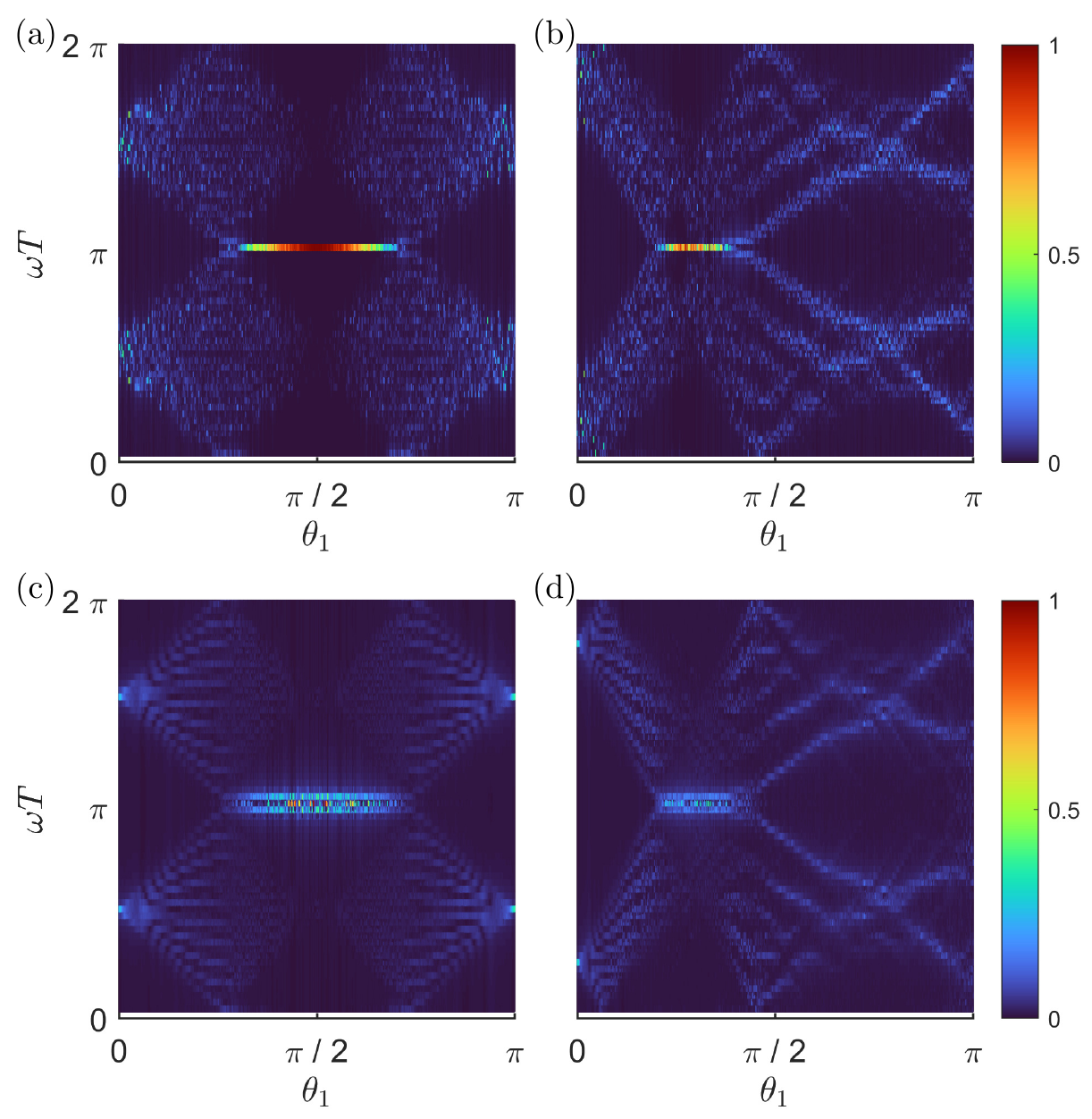}
\caption{(a-d) Frequency distributions of the response function for observable $\hat{X}_1$ and the color represents the amplitude of $A_X(\omega)$. In (a) and (b), $\sum_j \delta_j \hat{Z}_j$ is added which preserves the particle-hole symmetry, while in (c) and (d) $\sum_j \delta_j \hat{X}_j$ is added which breaks the particle-hole symmetry.
Parameters of (a, c) are the same as the black solid line in Fig.~\ref{fig:phase1}, while parameters of (b,d) are the same as the black solid line in Fig.~\ref{fig:phase2}. $\delta_j$ is randomly distributed in $[- 0.1 h_2^z, 0.1 h_2^z]$, which is only added in the second interval during a period. }
\label{fig:fredisdis}
\end{figure}

For D class or BDI class, the Hamiltonian obeys particle-hole symmetry, which means the Hamiltonian \eqref{Hcc} is invariant after particle-hole transformation $\hat{c}^\dagger_j\rightarrow(-1)^{j}\hat{c}_j$, $\hat{c}_j\rightarrow(-1)^{j}\hat{c}^\dagger_j$, and $i\rightarrow{-i}$. In the following, we add an operator $\sum_j \delta_j\hat{Z}_j$ as the particle-hole symmetry-preserving perturbation and $\sum \delta_j\hat{X}_j$ as the symmetry-breaking perturbation, with $\delta_j$ disordered parameters. As shown in Fig.~\ref{fig:fredisdis} (a) and (b), the frequency distributions of Eq.~(\ref{Oomega}), for D class and BDI class respectively, are almost unaffected by the symmetry-preserving perturbation. As a comparison, the sub-harmonic response is collapsed after adding symmetry-breaking perturbations for both classes. These results suggest that sub-harmonic response is robust to symmetry-preserving perturbations, which confirms the DTC is originated from topologically non-trivial phases.

\section{Summary}
\label{sec:sum}

We have systematically studied the Floquet time crystal in a periodically driven spin chain model by elaborating the topological classifications and phase diagrams after mapping it to a Majorana chain. In this paper, we have shown that the dynamics of the observable edge mode operator or end spin operator indeed exhibit robust sub-harmonic oscillation, which is a typical signature of the DTC. Besides, the rigidity and robustness of the DTC build on topologically non-trivial phases. Furthermore, since the system is a general spin chain model and can be strictly solved, it is potentially generalized to other interacting or dissipative systems. Our results might be helpful in deeply understanding the mechanism for other kinds of DTC. Besides, the model is easily implemented and the topological DTC for the end spin are readily realized in experiments.

{\bf Acknowledgements}
We would like to thank Hui Zhai, Wei Zheng, Chengshu Li, Yanting Cheng, Haifeng Tang and Fan Yang for helpful discussions. This work has been supported by the National Natural Science Foundation of China (Grant No. 12204428).

\appendix

\section{Proof of Eq.~(8)}
\label{app:a}
In this appendix, we prove that the effective Hamiltonian $\hat{H}_{\rm{eff}}$, defined by
\begin{equation}
\exp(-i\hat{H}_{{\rm eff}}T)=\exp(-i\hat{H_{2}}t_{2})\exp(-i\hat{H_{1}}t_{1}),
\end{equation}
is quadratic with $\hat{H}_{\rm{eff}}=\frac{1}{2}\hat{\Psi}^\dagger {H}_{\rm{eff}}\hat{\Psi}$, where $H_{\rm{eff}}$ has such form
\begin{equation}
    \exp \left(-i H_{\mathrm{eff}} T\right)=\exp \left(-i H_{2} t_{2}\right) \exp \left(-i H_{1} t_{1}\right).
\end{equation}

First, we rewrite $\hat{H}_{\rm{eff}}$ by means of the Baker–Campbell–Hausdorff formula as
\begin{equation}
\begin{aligned}
 &\hat{H}_{{\rm eff}}=\frac{1}{T}\{ \hat{H}_{1}t_{1}+\hat{H}_{2}t_{2}-\frac{it_{1}t_{2}}{2}[\hat{H}_{1},\hat{H}_{2}]\\
 &-\frac{t_{1}^{2}t_{2}}{12}[\hat{H}_{1},[\hat{H}_{1},\hat{H}_{2}]]-\frac{t_{1}^{2}t_{2}}{12}[\hat{H}_{2},[\hat{H}_{2},\hat{H}_{1}]]\ldots\}.
\label{HeffBCH}
\end{aligned}
\end{equation}

Second, since the right-hand of Eq.~(\ref{HeffBCH}) is made up of a series of commutators, we choose $[\hat{H}_1,\hat{H}_2]$ as a typical term to analyse. Because both $\hat{H}_1$ and $\hat{H}_2$ have the quadratic form as $\hat{H}_{m}=\frac{1}{2}\hat{\Psi}^\dagger H_m \hat{\Psi}$, we define a mapping from a matrix to an operator as $\hat{\Gamma}(H_m)=\frac{1}{2}\sum_{i,j}(H_m)_{i,j}\gamma_i\gamma_j$. 

Third, we prove the relation $\hat{\Gamma}([H_1,H_2])=[\hat{\Gamma}(H_1),\hat{\Gamma}(H_2)]$ in the following process,
\begin{widetext}
\begin{equation}
\begin{aligned}
\label{Eq:H1H2com}
    [\hat{H_{1}},\hat{H}_{2}]&=\frac{1}{4}\sum_{ij}\sum_{u,v}(H_{1})_{i,j}(H_{2})_{u,v}(\gamma_{i}\gamma_{j}\gamma_{u}\gamma_{v}-\gamma_{u}\gamma_{v}\gamma_{i}\gamma_{j})\\
    &=\frac{1}{4}\sum_{i,j}\sum_{u,v}(H_{1})_{i,j}(H_{2})_{u,v}\gamma_{i}(\delta_{u,j}-\gamma_{u}\gamma_{j})\gamma_{v}-\frac{1}{4}\sum_{i,j}\sum_{u,v}(H_{1})_{i,j}(H_{2})_{u,v}\gamma_{u}(\delta_{v,i}-\gamma_{i}\gamma_{v})\gamma_{j}\\
    &=\frac{1}{4}\sum_{i,j}\sum_{u,v}(H_{1})_{i,j}(H_{2})_{u,v}\delta_{u,j}\gamma_{i}\gamma_{v}-\frac{1}{4}\sum_{i,j}\sum_{u,v}(H_{2})_{u,v}(H_{1})_{i,j}\delta_{v,i}\gamma_{u}\gamma_{j}\\
    &-\frac{1}{4}\sum_{i,j}\sum_{u,v}(H_{1})_{i,j}(H_{2})_{u,v}(\delta_{u,i}-\gamma_{u}\gamma_{i})\gamma_{j}\gamma_{v}+\frac{1}{4}\sum_{i,j}\sum_{u,v}(H_{1})_{i,j}(H_{2})_{u,v}\gamma_{u}\gamma_{i}(\delta_{v,j}-\gamma_{j}\gamma_{v})\\
    &=\frac{1}{2}(H_{1}H_{2}-H_{2}H_{1})_{i,j}\gamma_{i}\gamma_{j}\\
    &=\hat{\Gamma}([H_1,H_2]),
\end{aligned}
\end{equation}
\end{widetext}
where we have used the property that matrices $H_1$ and $H_2$ are both
anti-symmetric. Using the relation of Eq.~\eqref{Eq:H1H2com}, we can prove $[\hat{H}_{1},[\hat{H}_{1},\hat{H}_{2}]]=\hat{\Gamma}([{H}_{1},[{H}_{1},{H}_{2}]])$, $[\hat{H}_{2},[\hat{H}_{2},\hat{H}_{1}]]=\hat{\Gamma}([{H}_{2},[{H}_{1},{H}_{2}]])$, and so on. In this way, every term in Eq.~\eqref{HeffBCH} can be rewritten as the quadratic form $\hat{\Gamma}$, so we have $\hat{H}_{\rm{eff}}=\hat{\Gamma}(H_{\rm{eff}})$, where
\begin{equation}
\begin{aligned}
&{H}_{{\rm eff}}=\frac{1}{T}\{ {H}_{1}t_{1}+\hat{H}_{2}t_{2}-\frac{it_{1}t_{2}}{2}[{H}_{1},{H}_{2}]\\
&-\frac{t_{1}^{2}t_{2}}{12}[{H}_{1},[{H}_{1},{H}_{2}]]-\frac{t_{1}^{2}t_{2}}{12}[{H}_{2},[{H}_{2},{H}_{1}]]\ldots\}.
\end{aligned}
\end{equation}
Again, we use the Baker–Campbell–Hausdorff formula and $H_{\rm{eff}}$ is obtained according to $\exp(-iH_{\rm{eff}}T)=\exp(-iH_1t_1)\exp(-iH_2t_2)$.

\section{Finite size effect}
\label{app:b}

\begin{figure}[t]
\includegraphics[width=8.3cm]{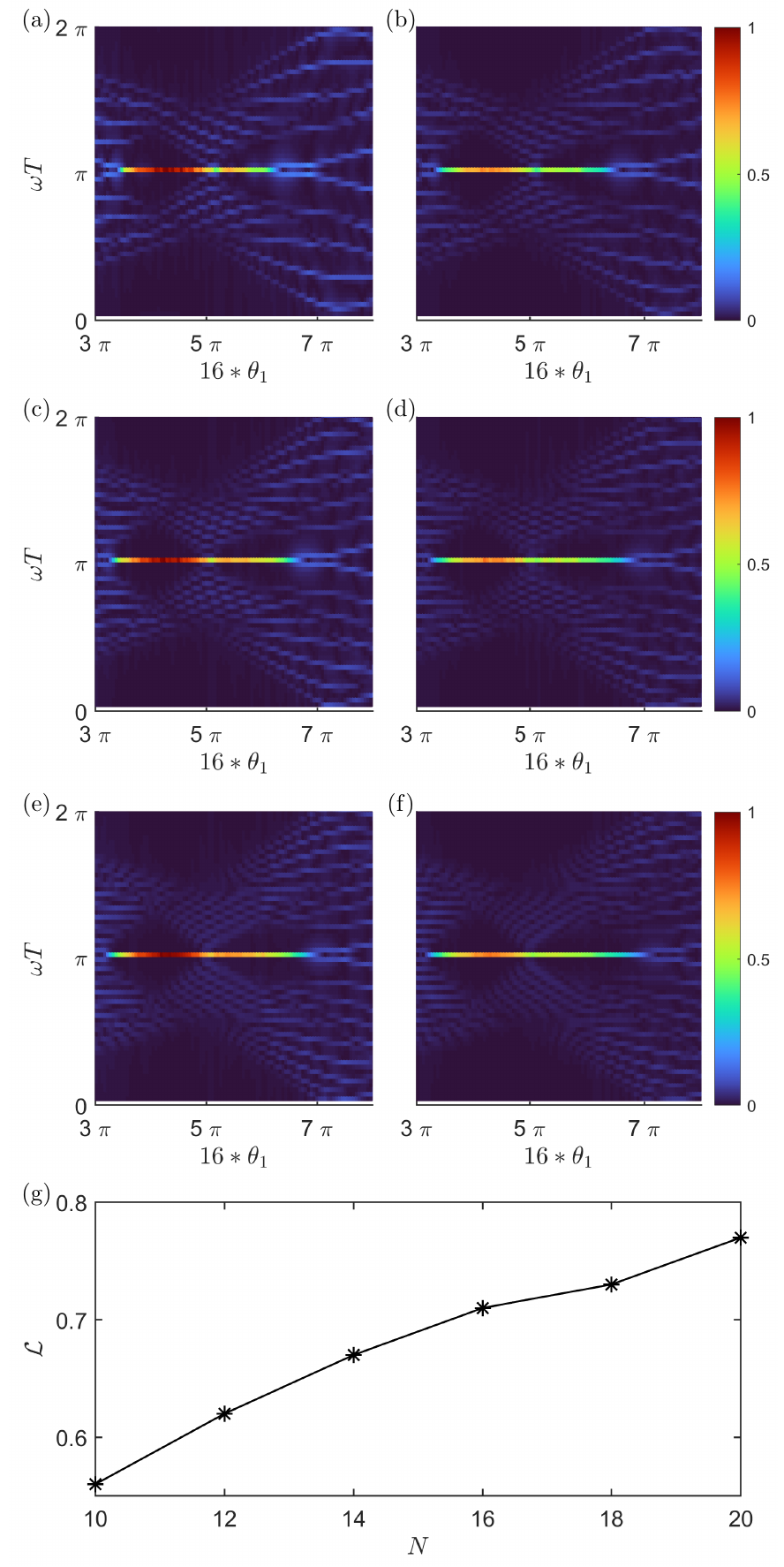}
\caption{(a-f) Frequency distributions of the response function for observable $\hat{X}_1$ and the color represents the amplitude of $A_X(\omega)$. The length of chian $N$ increases from $10$ to $12, 14, 16, 18, 20$, corresponding to (a-f), respectively. Parameters are the same as Fig.~\ref{fig:fredis} except for $\theta_1 \in [3 \pi / 16, \pi / 2]$, which is the topologically non-trivial regime with $\nu_\pi\neq 0$. (g) The regime of parameter $\theta_1$ satisfying $A_X(\omega = \pi / T) > 0.05$ varies with the length of chain $N$.}
\label{fig:fredisapp}
\end{figure}

In Fig.~\ref{fig:fredis}, the DTC regime, where $A_X(\omega)$ peaks at $\omega = \pi / T$, is smaller than the region of topologically non-trivial phases. We think this phenomenon is due to the finite size effect, which causes the quasi-energy of anomalous edge state is not perfectly $\pi / T$. As an example, we numerically calculate the DTC regimes for BDI class as the length of chain $N$ increases to show the finite size effect. The numerical results are shown in Fig.~\ref{fig:fredisapp}, from which we find the regime of parameter $\theta_1$ satisfying $A_X(\omega = \pi / T) > 0.05$ increases as the length of chain increases. It is inferred that the regime of parameter $\theta_1$ could match with the region of topologically non-trivial phases as $N\rightarrow\infty$.

\bibliography{ref.bib}

\begin{thebibliography}{47}
\bibitem [{foo()}]{foot1}%
  \BibitemOpen
  \href@noop {} {}\bibinfo {note} {The BdG constraint requires $C^{-1}HC=-H$
  with $C^2=1$, which means only Class D, BDI and DIII are allowed in 1-D
  system. However, for Class DIII it also requires $T^{-1}HT=-H$ with $T^2=-1$,
  which leads to a trivial result for a two-band model.}\BibitemShut {Stop}%
\end{thebibliography}



\end{document}